\documentclass[prl,twocolumn,aps,showpacs]{revtex4}
\usepackage{graphicx}
\usepackage{bm}
\usepackage{amssymb} 
\begin{document}
\title{Localization properties of random-mass Dirac fermions from real-space renormalization group}

\author{V. V. Mkhitaryan and M. E. Raikh}

\affiliation{Department of Physics, University of Utah, Salt Lake
City, UT 84112, USA}
\begin{abstract} Localization properties of random-mass Dirac
fermions for a realization of mass disorder, commonly referred to
as Cho-Fisher model, is studied on the D-class chiral network. We
show that a simple RG description captures accurately three
phases: thermal metal and two insulators with quantized Hall
conductances, as well as transitions between them (including
critical exponents). We find that, with no randomness in phases on
the links, transmission via the RG block exhibits a sizable
portion of {\it perfect resonances}. Delocalization occurs by
proliferation of these resonances to larger scales. Evolution of
the thermal conductance distribution towards metallic fixed point
is synchronized with evolution of {\it signs} of transmission
coefficients, so that delocalization is accompanied with {\it sign
percolation}.
\end{abstract}
 \maketitle

\noindent{\em Introduction.} As it was pointed out in Ref.
\cite{Grinstein}, there exists a profound connection between 2D
electron motion in random potential, $V(x,y)$, in a quantizing
magnetic field and the motion of 2D Dirac fermions with random
mass, $M(x,y)$, in a zero field. This connection can be traced to
the analogy of the drift of electron along closed equipotentials,
$V(x,y)=E$, and chiral motion of Dirac fermion along contours,
$M(x,y)=0$. Due to this similarity, conventional quantum Hall
transition, which takes place as equipotentials merge, has its
counterpart for Dirac fermions. However, critical behaviors for
these two transitions are different due to the different
underlying symmetries. In simple terms, the difference comes from
the fact that electron drifting along equipotential acquires a
random phase, while Dirac fermion moving along a line $M=0$ does
not. According to general classification \cite{AltZirn},
delocalization transition of Dirac fermions with mass randomness
belongs to the D-class \cite{Bocquet, HoChalker, ReadLudwig,
GruzReadLudwig, MerzChalker, SentFish, ReadGreen, BeenakkerCC08,
Beenakker08, BeenakkerR10, Beenakker10, EversChalker, EversMirlin,
Brouwer, Gruzberg05}: time-reversal and spin-rotational symmetries
are broken. It was realized that delocalization transition in the
D-class is related to criticality in the random-bond Ising model
\cite{ReadLudwig,GruzReadLudwig,MerzChalker}, thermal quantum Hall
effect in superconductors \cite{SentFish, ReadGreen, EversChalker,
EversMirlin, Brouwer, Gruzberg05} with either spin-orbit
scattering or with special type of pairing, and lately, to in-gap
transport in graphene with broken sublattice symmetry
\cite{BeenakkerCC08, Beenakker08, BeenakkerR10, Beenakker10,
Ziegler}.

Not only the critical exponent of localization radius for the
D-class is different from the quantum Hall critical exponent, but
the very picture of delocalization in the D-class is much less
intuitive than in the quantum Hall transition. This is because
transformation of the wave function corresponding to the drift
along $V(x,y)=E$ upon merging of two equipotentials is different
from transformation of chiral Dirac wave functions upon merging of
two contours $M(x,y)=0$. In both cases, the reflection and
transmission coefficients, $r$ and $t$, at the point of merging
can be chosen real with $r^2+t^2=1$, ensuring the current
conservation. In the quantum Hall case, the {\it signs} of $r$ and
$t$ can be chosen arbitrary, since change of the sign can be
absorbed into the random drift-phase. By contrast, there is no
such freedom in the D-class. Indeed, the phase accumulated by
Dirac edge state along a contour $M(x,y)=0$ is always equal to
$\pi$, which is the consequence of the pseudospin structure of the
eigenfunction \cite{Grinstein}. If scattering matrices
\begin{equation}
\label{smat} S_i=\left(\begin{array}{cc}
t_i&r_i\\
r_i&-t_i
\end{array}\right)
\end{equation}
are the same for all points of contact $i$ (without any randomness
in signs), it is obvious that delocalization will take place at
$r=t=1/\sqrt{2}$, which corresponds to $\langle M(x,y)\rangle=0$.
Then the critical exponent is $\nu=1$, which is a consequence of
in-gap tunneling of the Dirac fermion.
\begin{figure}[t]
\vspace{-0.4cm}
\centerline{\includegraphics[width=100mm,angle=0,clip]{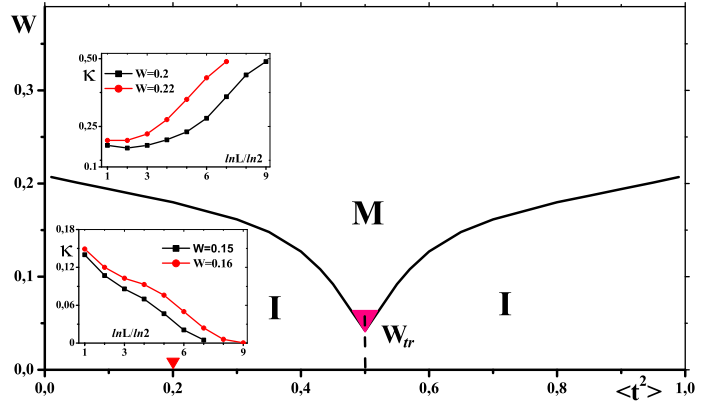}}
\vspace{-0.6cm} \caption{(Color online) CF model. The boundary on
the plane gap-disorder $(\langle t^2\rangle, W)$ (solid line)
separates metallic and insulating fixed points of the RG
transformation Eq. (\ref{RGstep}). Dashed line-the boundary
between insulators with different values of quantized thermal Hall
conductance. Purple triangle marks a tricritical point at
$W_{tr}\approx 0.06$. Upper and lower insets show the evolution of
the portion, $\kappa$, of negative values of reflection
coefficients, $r_i$, with the sample size calculated for $\langle
t^2\rangle=0.2$ in metallic and insulating phases, respectively. }
\label{phasediag}
\end{figure}

An amazing property of the D-class is that disorder in {\it signs}
of $r_i$ and $t_i$ has a drastic effect on delocalization
transition \cite{signfootnote}. Moreover, for a given degree of
the net sign disorder, a particular way in which it is introduced
\cite{ReadLudwig, GruzReadLudwig, ReadGreen, ChoFisher,
KagaChalker} can turn the system either into metal (M) or into
insulator (I). In particular, if disorder is introduced by
randomly changing signs of {\it columns} in Eq. (\ref{smat}) with
probability, $W$, then {\it arbitrarily small} $W$ leads to
delocalization at any average mass, $\langle M(x,y)\rangle\propto
\langle t_i^2\rangle-1/2$. This fact was established numerically
in Ref. \cite{KagaChalker}, where the corresponding randomness was
dubbed $O(1)$. For two other types of sign disorder: random-bond
Ising model \cite{ReadLudwig,GruzReadLudwig,MerzChalker} and
Cho-Fisher (CF) model \cite{ChoFisher}, the metallic phase is
either absent or emerges when $W$ exceeds certain $W_c$,
correspondingly. The above two facts were also established
numerically by studying transmission of finite-width (up to ${\bf
M}=256$) stripes \cite{BeenakkerR10, KagaChalker, KagaNemirPRL,
KagaNemir}.
\begin{figure}[t]
\centerline{\includegraphics[width=85mm,angle=0,clip]{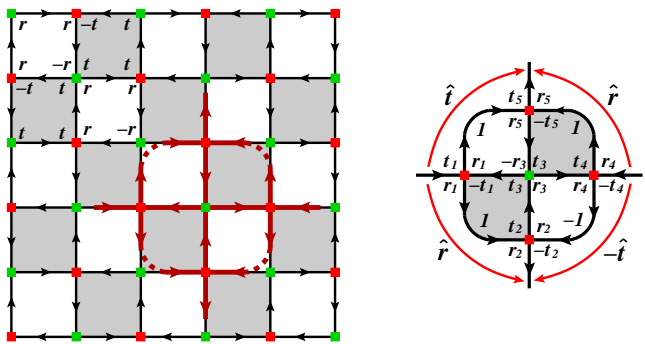}}
\caption{(Color online) Left: Chiral network for the CF model.
Sign arrangement of $t$ and $r$ at the nodes ensures $\pi$ flux
per plaquette \cite{EversChalker}. Red lines single out a group of
five nodes constituting an RG block. Right: Illustration of the RG
transformation Eqs. (\ref{RGstep}), (\ref{rRGstep}) for elements
of renormalized node scattering matrix on the red sublattice.
Truncation procedure (dashed lines on the original lattice)
enforces, via sign structure in the nodes, the factors $1$ and
$-1$ in the corners. These factors are taken into account in
Eqs.(\ref{RGstep}), (\ref{rRGstep}). } \label{block}
\end{figure}

In conventional quantum Hall effect, emergence of an isolated
delocalized state at $\langle t_i^2\rangle=1/2$ is qualitatively
transparent, since the Chalker-Coddington model \cite{CC} is a
quantum version of a classical percolation, which possesses
self-duality. By contrast, there is no classical version of the
D-class. Thus there is no qualitative explanation how a particular
local sign disorder in $t_i$ and $r_i$ results either in metal or
in insulator at large distances. Moreover, the fact that critical
disorder  in CF model is weakly sensitive to $\langle M\rangle$
\cite{KagaChalker, KagaNemirPRL} seems counterintuitive. Indeed,
it implies that delocalization takes place no matter how weakly
the $M=0$ contours are coupled to each other.

In this paper we demonstrate that above puzzles of the D-class
delocalization find a natural explanation within a simple
real-space renormalization group (RG) description. This
description is developed below by adjusting to the D-class the
approach of Ref. \cite{aram} developed for conventional quantum
Hall effect. We focus on the CF model most general in the sense
that its phase diagram in the $\bigl(\langle t_i^2\rangle,
W\bigr)$-plane contains all three (metal and two quantum Hall
insulator) phases.

The RG procedure \cite{aram} prescribes how the magnitude and {\it
sign} of effective transmission coefficient, $\hat{t}$, evolve
with the system size, $L$. Within RG we demonstrate that I-M
transition (finite $\hat{t}^2$ at large $L$) occurs { \it at the
same} $W_c$ where the distribution of the {\it amplitude},
$\hat{t}$, becomes {\it symmetric} with respect to $\hat{t}=0$.
For smaller $W$ (insulating phase), the initial signs of $t_i$ are
"forgotten" with increasing $L$. In this sense, the I-M transition
can be viewed as {\it sign percolation}. Remarkably, RG
description appears to be very accurate on the quantitative level.
In fact, it reproduces the entire phase diagram found in
\cite{KagaChalker, KagaNemirPRL, KagaNemir}. In addition, we find
the following universal distribution of the conductance,
$G=\hat{t}^2$, of the thermal metal
\begin{equation}
\label{distribution} P(G)=0.414[G(1-G)]^{-0.6}.
\end{equation}
Within RG, insensitivity to $\langle M\rangle$ emerges as a result
of high (resonant) transmission via certain local disorder
configurations; importantly, these resonances proliferate to large
distances.

\noindent{\em RG transformation.} In the "clean" case, the network
model of D-class is shown in Fig. \ref{block}. The signs of
transmission and reflection at each node ensure the $\pi$-flux per
plaquette. With respect to signs, scattering matrices of the nodes
are different for two sublattices of the square lattice, so that
$S$-matrix has the form Eq. (\ref{smat}) for red nodes and
\begin{equation}
\label{Sblue} S^\prime_i=\left(\begin{array}{cc}
t_i&-r_i\\
r_i&t_i
\end{array}\right)
\end{equation}
for green nodes. The RG superblock shown in Fig. \ref{block}
consists of four red and one green nodes. With signs in the
$S$-matrices chosen accordance with Fig. \ref{block}, scattering
matrix of this superblock reproduces the form Eq. (\ref{smat})
\cite{footnote} with effective transmission coefficient given by
\begin{equation}
\label{RGstep} \hat{t} = \frac{t_1t_5 (r_2r_3r_4 + 1) + t_2t_4
(r_1r_3r_5 + 1) + t_3(t_1t_4 + t_2t_5)}{(r_3 + r_1r_5) (r_3 +
r_2r_4) + (t_3 + t_1t_2) (t_3 + t_4t_5)}.
\end{equation}
At the next RG step, values of $\hat{t}_i$ generated by Eq.
(\ref{RGstep}), {\it including the signs}, should be placed into
red nodes of renormalized lattice, which has a doubled lattice
constant \cite{Reynolds}.

\noindent{\em The limit $W=0$.} As a first test of applicability
of the RG description we apply Eq. (\ref{RGstep}) to an ordered
system. Upon setting all $t_i$ equal to $t$ and all $r_i$ equal to
$\sqrt{1-t^2}$, we see that $\hat{t}=t=1/\sqrt{2}$ is indeed a
fixed point, corresponding to zero mass. Finite $|t^2-1/2|$ plays
the role of a finite mass, $M$, in a clean system. From Dirac
equation it follows that transmission disappears at lengths
exceeding tunneling length $1/M$. This means that the $W=0$
critical exponent is $\nu=1$. On the other hand, Eq.
(\ref{RGstep}) transforms finite $|t^2-1/2|$ into
$|\hat{t}^2-1/2|=\tau|t^2-1/2|$, where
\begin{equation}
\label{coeff}
\tau=\sum_{i=1}^5c_i=\sum_{i=1}^5\frac{\partial\hat{t}^2}{\partial
t^2}{\Big |} _{t^2=1/2}.
\end{equation}
Elementary calculation yields $c_1=c_2=c_4=c_5=\sqrt{2}-1$,
$c_3=3-2\sqrt{2}$, so that $\tau=2\sqrt{2}-1$. Since $\tau>1$,
transmission disappears upon $n$ subsequent steps, where $n$ is
determined by the condition, $|t^2-1/2|\tau^n\sim 1$. For such $n$
the unit cell is $\xi=2^n$ and should be identified with the
in-gap tunneling length. From the above two conditions we find
that $\xi\sim|t^2-1/2|^{-\nu}=|M|^{-\nu}$, where $\nu=\ln2/\ln
\tau\approx 1.15$, {\it i.e.,} the RG value of the exponent is
only $15\%$ different from the exact value.

Before introducing disorder in signs of $t_i$ and $r_i$, we
elucidate another property of $t^2=1/2$ fixed point: initial {\it
symmetric} distribution of $t^2$ around $1/2$ not only remains
symmetric but {\it narrows} upon renormalization. Indeed, the
width of distribution after one RG step evolves as $\langle
\delta\hat{t}^2\rangle = \sum_{i=1}^5c_i^2\langle \delta
t_i^2\rangle\approx0.7 \langle \delta t_i^2\rangle$. This means
that the critical states at the above fixed point do not exhibit
mesoscopic fluctuations, in contrast to conventional quantum Hall
critical state. We will see below that thermal metal does exhibit
strong mesoscopic fluctuations.
\begin{figure}[t]
\vspace{-0.6cm}
\centerline{\includegraphics[width=100mm,angle=0,clip]{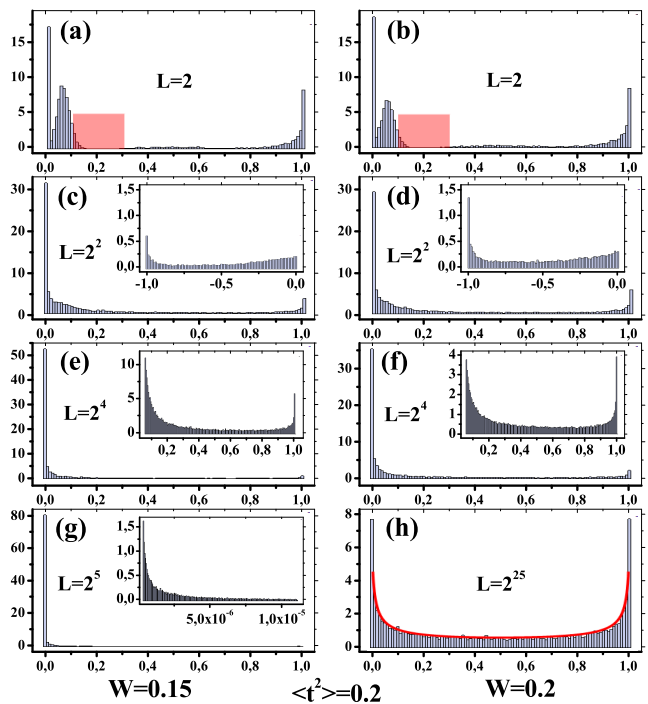}}
\vspace{-0.8cm} \caption{(Color online)  RG evolution with system
size, $L$, of the initial distribution of $t_i^2$ (purple box) is
shown for two disorder magnitudes, $W=0.15$ (left) and $W=0.2$
(right). Insets: (c) and (d) Distribution of the {\it amplitude}
reflection coefficient, $r_i$, in the domain $-1<r_i<0$;  (e) and
(f) Distribution of $t_i^2$ with $t_i^2<0.025$ removed illustrate
precursors of insulator and metal, respectively; (g) Distribution
of $t_i^2$ in the domain, $10^{-4}<t_i^2<10^{-5}$ illustrates
approach to the insulator.} \label{evolution}
\end{figure}

\noindent{\em Resonances.} Suppose that in Eq. (\ref{RGstep}) all
$t_i$ are small. Corresponding $r_i$ are close to 1. It might seem
that $\hat{t}$ is even smaller, $\propto t_i^2$. At this point we
divulge the following remarkable property of the transformation
Eq. (\ref{RGstep}). Let us choose the following sign combination:
\begin{equation}
\label{resonance} r_1=r_2=r_3=r_4=r, \quad r_5=-r.
\end{equation}
Substituting the above values into Eq. (\ref{RGstep}), we find
that $\hat{t}$ is {\it identically} equal to $\pm 1$, {\it i.e.,}
{\it resonant tunneling} takes place. It might also seem that such
resonant configurations are irrelevant due to vanishing
statistical weight. On the contrary, we will see that these
resonances play a key role in delocalization, since they persist
even when $r_i$ are not equal. Evidence to this fact can be found
in Fig. \ref{evolution}a,b. Bare values of $t_i^2$ are
homogeneously distributed in the interval, $0.1<t_i^2<0.3$. For a
modest randomness in signs, Eq. (\ref{RGstep}) transforms the
box-like distribution into a three-peak distribution; the left
peak corresponds to resonant reflection, the middle peak
corresponds to expected overall reduction of transmission, while
the right peak comes from resonant transmission. We see that, with
spread in the bare $t_i^2$ values, the portion of resonant
transmission is considerable, about $8\%$.
Figs. \ref{evolution}c-h illustrate that the competition between
the resonant transmission peak and low-$t^2_i$ peaks decides
whether the system evolves into a metal or an insulator. In the
metallic phase, Figs. \ref{evolution}d, \ref{evolution}f, and
\ref{evolution}h, the resonant transmission peak continuously
broadens and grows until the distribution becomes symmetric around
$t^2=1/2$. In the insulating phase, Figs. \ref{evolution}c,
\ref{evolution}e, and \ref{evolution}g, the resonant tunneling
peak gets gradually suppressed with the system size, and
large-scale distribution flows to $t_i^2\rightarrow 0$. Details of
determination of the I-M boundary are described below.

\noindent{\em CF sign disorder.} According to prescription of Ref.
\cite{EversChalker} for the isotropic version of the CF model, for
a given positive $t$ the values of $t_i$ are chosen to be
\begin{equation}
\label{CFtd} t_i=\left\{
\begin{array}{l} t,\quad\text{with probability}\quad
\left(1-\frac W2\right),\\
\,\\
-t,\quad\text{with probability}\quad \frac W2.
\end{array}
\right.
\end{equation}
Correspondingly,
\begin{equation}
\label{CFrd} r_i=\left\{
\begin{array}{l} \sqrt{1-t^2},\quad\text{with probability}\quad
\left(1-\frac W2\right),\\
\,\\
-\sqrt{1-t^2},\quad\text{with probability}\quad \frac W2.
\end{array}
\right.
\end{equation}
The RG transformation for $r_i$ has a form
\begin{equation}
\label{rRGstep} \hat{r} = \frac{r_1r_2 (t_3t_4t_5 + 1) + r_4r_5
(t_1t_2t_3 + 1) + r_3(r_1r_4 + r_2r_5)}{(r_3 + r_1r_5) (r_3 +
r_2r_4) +(t_3 + t_1t_2) (t_3 + t_4t_5)},
\end{equation}
which is in accord with Eq. (\ref{RGstep}). With prescription Eqs.
(\ref{CFtd}), (\ref{CFrd}), and Eqs. (\ref{RGstep}),
(\ref{rRGstep}), Mathematica performs each subsequent RG step
almost instantaneously even for a dense sample of $10^6$ $t_i$ and
$r_i$. We started with checking that for initial box-like
distribution of $t^2$ in the interval, $t^2_{min}<t^2<t^2_{max}$,
the RG evolution depends only on the average, $\langle
t^2\rangle=(t^2_{min}+t^2_{max})/2$, but not on the width,
$(t^2_{max}-t^2_{min})$. For particular box, $t^2_{min}=0.1$,
$t^2_{max}=0.3$, results shown in Fig. \ref{evolution} illustrate
that distinction between two disorder values, $W=0.15$ and
$W=0.2$, emerges at system size, $L=2^5$. At the same time, the
difference in {\it signs distribution} of $r_i$ develops already
at the second step. This is illustrated in Figs.
\ref{evolution}c,d insets. This behavior is generic: memory about
initial sign disorder in $r_i$ is quickly forgotten in the
insulating phase (all $r_i$ have are positive after 3-4 steps),
whereas in the metallic phase both signs of $r_i$ are completely
equilibrated after 3-4 steps. Nevertheless, complete convergence
to the symmetric fixed point distribution $P(t^2)$ given by Eq.
(\ref{distribution}) takes place at very large size, $L=2^{25}$.
\begin{figure}[t]
\vspace{-0.6cm}
\centerline{\includegraphics[width=100mm,angle=0,clip]{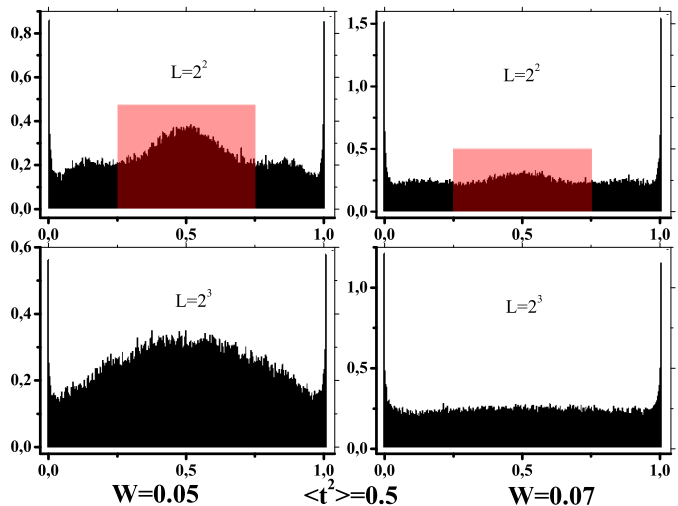}}
\vspace{-0.8cm}\caption{(Color online)  RG evolution with sample
size, $L$, of the symmetric distribution of $0.25<t_i^2<0.75$
(purple box) is shown for two magnitudes of disorder, $W=0.05$
(left) and $W=0.07$ (right). } \label{central}
\end{figure}

RG evolutions depicted in Fig. \ref{evolution} suggest that a M-I
transition point lies within $0.15<W_c<0.2$. By gradually
shrinking this interval, one can find $W_c$ with high accuracy.
The results for different $\langle t^2\rangle$, which are in
general agreement with numerics of Refs. \cite{KagaChalker,
KagaNemirPRL, KagaNemir}, are shown in Fig. \ref{phasediag}. We
see that the I-M boundary is approximately horizontal, except for
the interval, $0.4<\langle t^2\rangle<0.6$. Within this interval
the boundary rapidly drops to the value, $W_{tr}=0.06$, which we
identify with tricritical point. Insets in Fig. \ref{phasediag}
illustrate our statement concerning sign percolation. They show
how the portion, $\kappa$, of negative $r_i$-values Eq.
(\ref{rRGstep}) evolves with the sample size. Equilibration of
signs in metal, $W>W_c$, implies that $\kappa$ grows towards
$\kappa=1/2$. We see that the bigger is $(W-W_c)$ the faster is
the growth. By contrast, a decrease of $\kappa$ with $L$ in
insulator implies that all $r_i$ become positive after several
steps. It is seen that erasing of signs is more efficient at
smaller $W$. As signs are erased, the magnitude, $\langle
r^2\rangle$ approaches $1$. We used the rate of this approach to
estimate the critical exponent of the I-M transition. Assuming
that $(1- \langle r^2(L)\rangle)$ is a function of a single
parameter, $(W_c-W)^{\nu}L$, we found $\nu\approx 1.2$ for initial
distribution with $\langle t^2\rangle=0.2$ (red triangle mark in
Fig. \ref{phasediag}).

\noindent{\em Tricritical point.} In the RG language, trictritical
point on the vertical axis, $t^2=1/2$, suggests that a symmetric
initial distribution of $t_i^2$ evolves to
$P(t^2)=\delta(t^2-1/2)$ for $W<W_{tr}$, and to metallic fixed
point, Eq. (\ref{distribution}) for $W>W_{tr}$. Fig.
(\ref{central}) shows the evolution of symmetric box-like
distribution $0.25<t_i^2<0.75$ for two $W$-values. We see that for
$W=0.05$ and $W=0.07$ these evolutions are different. For $W=0.07$
the distribution at $L=2^3$ is flat. After a few steps the
histogram bends down at the center and distribution flows towards
metallic fixed point Eq. (\ref{distribution}). For $W=0.05$, the
box-like initial distribution develops a maximum at $L=2^3$, Fig.
\ref{central}c. This narrowing suggests a flow towards the ordered
fixed point, $t=r=1/\sqrt{2}$. By gradually shrinking the
$W$-interval we locate the value $W_{tr}\approx 0.06$, which
separates the behavior Fig. \ref{central}c (maximum at the center)
and Fig. \ref{central}d (minimum at the center). We identify
$W_{tr}$ with tricritical point \cite{KagaChalker, EversChalker,
KagaNemirPRL, KagaNemir, Beenakker10}. The smaller is $W$ the more
pronounced is the shrinking of initial distribution with $L$.
Note, however, that this shrinking eventually stops: at large $L$
maximum at the center is accompanied by satellite peaks at $t^2=0$
and $t^2=1$, which gradually take over and drive the system to
metal. Similar complications were pointed out in Ref.
\cite{EversChalker}. The behavior of sings of $t_i$ ($r_i$) is
synchronized with distribution of $t_i^2$ ($r_i^2$). Namely, as
the distribution shrinks, the portion, $\kappa$, changes from $W$
down monotonically. As the distribution turns back to metal,
$\kappa$ starts to grow towards $\kappa=1/2$.

\noindent{\em Discussion} The value, $W_c\approx 0.2$, and the
fact that it depends weakly on $t_i^2$, can be inferred from the
calculation of likelihood, ${\cal P}_W$, of resonant
configurations of the type Eq. (\ref{resonance}).  These
configurations occur when denominator in Eq. (\ref{RGstep}) is
small.  When all $t_i$ are small and $|r_i|$ close to $1$, this
condition requires that the product, $(r_3+r_1r_5)(r_3+r_2r_4)$,
is small. On the other hand, if both brackets are small, we will
have $\hat{t}\sim t_i$, {\it i.e.,} the resonance is absent (due
to suppression of the numerator). The probability that only one of
the brackets is small is given by
\begin{equation}
\label{likehood} {\cal
P}_W=4W(1-W)^3+4W^3(1-W)=\frac{1-(1-2W)^4}2.
\end{equation}
It turns out that ${\cal P}_W$ is almost flat and close to $1/2$
in the wide interval, $0.2<W<0.8$. This is the consequence of the
fact that first three derivatives of ${\cal P}_W$ are all zero at
$W=1/2$. Therefore if the resonances do not proliferate at
$W\approx 0.2$, they will not proliferate upon further increase of
$W$. This explains why, at small $t_i$, the critical $W_c$ is
close to $0.2$. Another feature of the phase diagram Fig.
\ref{phasediag}, weak dependence of $W_c$ on $\langle t^2\rangle$,
can be traced to the fact that for resonant configuration Eq.
(\ref{resonance}) we have $\hat{t}^2=1$ regardless of the value
 of $r$.

\noindent{\em Acknowledgments.} We are grateful to I. Gruzberg, V.
Kagalovsky, and
J. T. Chalker for motivation. M. E. R. is grateful KITP Santa
Barbara for hospitality. This work was supported by the BSF grant
No. 2006201 and DOE Grant No. DE-FG02-06ER46313.


\begin{references}
\bibitem{signfootnote} In the language of Ref.
\cite{Beenakker10} (see also Refs. \cite{GurRadz, BeenakkerPRL10})
the randomness in signs is related to the small-scale behavior
(vorticity) near the saddle point in $M(x,y)$.
\end{references}
\end{document}